\documentclass[prd,showpacs,amsmath,amssymb,superscriptaddress,nofootinbib]{revtex4}

\newcommand{\cqg}{Classical Quantum Gravity\ }

\begin{document}

\title{Non-minimal monopoles of the Dirac type as realization of the censorship conjecture}

\author{Alexander B. Balakin}
\email{Alexander.Balakin@ksu.ru} \affiliation{Department of
General Relativity and Gravitation, Kazan State University,
Kremlevskaya str. 18, Kazan 420008, Russia}
\author{Heinz Dehnen}
\email{Heinz.Dehnen@uni-konstanz.de} \affiliation{Universit\"at
Konstanz, Fachbereich Physik, Fach M 677, D-78457, Konstanz,
Germany}
\author{Alexei E. Zayats}%
\email{Alexei.Zayats@ksu.ru} \affiliation{Department of General
Relativity and Gravitation, Kazan State University, Kremlevskaya
str. 18, Kazan 420008,
Russia}%

\begin{abstract}
We discuss a class of exact solutions of a three-parameter
non-minimally extended Einstein-Maxwell model, which are
attributed to non-minimal magnetic monopoles of the Dirac type. We
focus on the investigation of the gravitational field of Dirac
monopoles for those models, for which the singularity at the
central point is hidden inside of an event horizon independently
on the mass and charge of the object. We obtained the
relationships between the non-minimal coupling constants, for
which this requirement is satisfied. As explicit examples, we
consider in detail two one-parameter models: first, non-minimally
extended Reissner-Nordstr\"om model (for the magnetically charged
monopole), second, the Drummond-Hathrell model.
\end{abstract}


\maketitle

\section{Introduction}\noindent

In 1969 R.~Penrose formulated the so-called cosmic censorship
conjecture \cite{Penrose}, which assumes, in particular, that
singularities have to be hidden inside of an event horizon and
invisible to distant observers \cite{Townsend,Wald}. In the
minimal Einstein theory there exists a number of exact solutions,
which can be considered as counterexamples to this censorship
conjecture. For instance, the static spherically symmetric
solutions to the Einstein equations with massless scalar field
\cite{Wyman} {\it always}\/ describe a naked singularity
\cite{Roberts,Dehnen93}. Naked singularities also appear, when we
deal with the Reissner-Nordstr\"om metric, if
$M^2<Q_{(e)}^2+Q_{(m)}^2$ ($M$, $Q_{(e)}$, $Q_{(m)}$ are the mass,
electric and magnetic charges, respectively), or with the Kerr
metric, if $M<|J|$ ($J$ is an angular momentum). The solution for
individual electron with $M\ll |Q_{(e)}|$ (in the geometrical
units) gives the simplest example of the naked singularity,
because the gravitational attraction is negligible compared to the
Coulomb repulsion, and the corresponding metric has no horizons.

We assume, that a {\it non-minimal} interaction between
electromagnetic and gravitational fields can eliminate this
contradiction, i.e., the non-minimality results in the appearance
of a new horizon, which hides the singular central point. Indeed,
curvature coupling constants, which are involved into the
non-minimal three-parameter Einstein-Maxwell model, can be
naturally associated with characteristic lengths of the
non-minimal interaction and thus, at least one extra parameter,
$r_q$, appears (see, e.g., \cite{BL05,BBL}) in addition to the
standard Schwarzschild radius $r_g$ and Reissner-Nordstr\"om
radius $r_Q$. This non-minimal extension sophisticates essentially
the causal structure of space-time around the charged objects, and
the appearance of an additional horizon, related to the censorship
conjecture, becomes possible.

In order to illustrate this idea, we consider now exact solutions
of the non-minimal Einstein-Maxwell model describing the magnetic
monopoles of the Dirac type. In the minimal theory the solution of
this type demonstrates a naked singularity in the center,
nevertheless, the curvature coupling is shown to lead to the
hiding of this singularity inside of the non-minimal horizon. The
exact three-parameter non-minimal solutions of the Dirac type can
be represented in an explicit analytic form, which simplifies the
discussion. These solutions can be considered as a direct
reduction of the solutions, obtained for the non-minimal $SU(2)$
symmetric quasi-Abelian Wu-Yang monopole \cite{BaZaPLB}, to the
model with $U(1)$-symmetry.

The paper is organized as follows. In Section \ref{G2} we discuss shortly
the fundamentals of the model and represent a three-parameter
family of exact solutions describing non-minimal Dirac monopole.
In Section~\ref{numbersection} we consider relationships between
three coupling constants, for which the space-time metric
possesses a singularity ``clothed'' in horizon for {\it arbitrary}
mass and charge of the object. In Subsection~\ref{RNsection} we consider
non-minimal horizons for the exactly integrable model of the
Reissner-Nordstr\"om type. In Subsection~\ref{DHsection} we discuss
in detail the one-parameter Drummond-Hathrell model, the horizon
radius being obtained and estimated explicitly. In the last
Section we summarize the results.

\section{Three-parameter family of exact solutions for non-minimal
monopoles of the Dirac type}\label{G2}

\subsection{Non-minimally extended Einstein-Maxwell theory}\noindent

The three-parameter non-minimal Einstein-Maxwell theory can be
formulated in terms of the action functional
\begin{equation}
S_{{\rm NMEM}} = \int d^4 x \sqrt{-g}\
\left[\frac{R}{8\pi}+\frac{1}{2}F_{ik} F^{ik}+\frac{1}{2}
{\cal R}^{ikmn}F_{ik} F_{mn}\right]\,.\label{act}%
\end{equation}%
Here $g = {\rm det}(g_{ik})$ is the determinant of the metric
tensor $g_{ik}$, $R$ is the Ricci scalar. The Latin indices
without parentheses run from 0 to 3. The Maxwell tensor $F_{ik}$
is expressed, as usual, in terms of a potential four-vector $A_k$
\begin{equation}
F_{ik} = \nabla_i A_k - \nabla_k A_i \,, \label{F}%
\end{equation}
where the symbol $\nabla_i$ denotes the covariant derivative. The
tensor ${\cal R}^{ikmn}$ is defined as follows (see \cite{BL05}):
\begin{equation}
{\cal R}^{ikmn} \equiv \frac{q_1}{2}R\,(g^{im}g^{kn}-g^{in}g^{km})
+ \frac{q_2}{2}(R^{im}g^{kn} - R^{in}g^{km} + R^{kn}g^{im}
-R^{km}g^{in}) + q_3 R^{ikmn} \,, \label{sus}
\end{equation}%
where $R^{ik}$ and $R^{ikmn}$ are the Ricci and Riemann tensors,
respectively, and $q_1$, $q_2$, $q_3$ are the
phenome\-no\-logi\-cal parameters describing the non-minimal
coupling of electromagnetic and gravitational fields. The
variation of the action functional with respect to potential $A_i$
yields
\begin{equation}
\nabla_k \left( F^{ik} + {\cal R}^{ikmn}F_{mn}\right)=0 \,.
\label{HikR}
\end{equation}
In a similar manner, the variation of the action with respect to
the metric yields
\begin{equation}
R_{ik} - \frac{1}{2} R \ g_{ik} = 8\pi\,T^{({\rm eff})}_{ik} \,.
\label{Ein}
\end{equation}
The effective stress-energy tensor $T^{({\rm eff})}_{ik}$ can be
divided into four parts:
\begin{equation}
T^{({\rm eff})}_{ik} =  T^{(M)}_{ik} + q_1 T^{(I)}_{ik} + q_2
T^{(II)}_{ik} + q_3 T^{(III)}_{ik} \,. \label{Tdecomp}
\end{equation}
The first term $T^{(M)}_{ik}$:
\begin{equation}
T^{(M)}_{ik} \equiv \frac{1}{4} g_{ik} F_{mn}F^{mn} -
F_{in}F_{k}^{\ n} \,, \label{TYM}
\end{equation}
is a stress-energy tensor of the pure electromagnetic field. The
definitions of other three tensors are related to the
corresponding coupling constants $q_1$, $q_2$, $q_3$:
\begin{equation}%
T^{(I)}_{ik} = R\,T^{(M)}_{ik} -  \frac{1}{2} R_{ik} F_{mn}F^{mn}
+ \frac{1}{2} \left[ \nabla_{i} \nabla_{k} - g_{ik} \nabla^l
\nabla_l \right] \left[F_{mn}F^{mn} \right] \,, \label{TI}
\end{equation}%

\[%
T^{(II)}_{ik} = -\frac{1}{2}g_{ik}\biggl[\nabla_{m}
\nabla_{l}\left(F^{mn}F^{l}_{\ n}\right)-R_{lm}F^{mn} F^{l}_{\ n}
\biggr] - F^{ln}
\left(R_{il}F_{kn} + R_{kl}F_{in}\right)-R^{mn}F_{im} F_{kn} {} \]%
\begin{equation}%
{} - \frac{1}{2} \nabla^m \nabla_m \left(F_{in} F_{k}^{ \
n}\right)+\frac{1}{2}\nabla_l \left[ \nabla_i \left( F_{kn}F^{ln}
\right) + \nabla_k \left(F_{in}F^{ln} \right) \right] \,,
\label{TII}
\end{equation}%

\begin{equation}%
T^{(III)}_{ik} = \frac{1}{4}g_{ik} R^{mnls}F_{mn}F_{ls}-
\frac{3}{4} F^{ls} \left(F_{i}^{\ n} R_{knls} + F_{k}^{\
n}R_{inls}\right) - \frac{1}{2}\nabla_{m} \nabla_{n} \left[
F_{i}^{ \ n}F_{k}^{ \ m} + F_{k}^{ \ n} F_{i}^{ \ m} \right] \,.
\label{TIII}
\end{equation}%
One may check directly that the tensor $T^{({\rm eff})}_{ik}$
satisfies the equation $\nabla^k T^{({\rm eff})}_{ik} =0$.

Below we consider non-minimally extended Einstein-Maxwell
equations (\ref{HikR}), (\ref{Ein})-(\ref{TIII}) for the case of
the static spherically symmetric space-time metric
\begin{equation}\label{metrica}
ds^2=\sigma^2Ndt^2-\frac{dr^2}{N}-r^2 \left( d\theta^2 +
\sin^2\theta d\varphi^2 \right) \,,
\end{equation}
where $N$ and $\sigma$ are functions of the radial variable $r$
only.

\subsection{Minimal solution with naked singularity as a starting point}

In the minimal Einstein-Maxwell theory the exact static
spherically symmetric solution of the Reissner-Nordstr\"om type is
the following
\begin{equation}\label{RN}
\sigma(r) = 1 \,, \quad N(r) = 1 - \frac{2M}{r} +
\frac{Q^2_{(e)}+Q^2_{(m)}}{r^2} \,.
\end{equation}
When $M<\sqrt{Q^2_{(e)}+Q^2_{(m)}}$, there are no horizons, and
the central point $r=0$ is classified as the naked singularity.
When $Q_{(e)}=0$ and $M< |Q_{(m)}|$, one deals with a magnetic
naked singularity.

The non-minimal Einstein-Maxwell model for the static spherically
symmetric space-time and central electric and magnetic  charges
was studied for two special sets of the coupling constants, the
first one satisfies the equalities $q_1+q_2+q_3=0$ and
$2q_1+q_2=0$ (see, e.g., \cite{HornPRD,HornJMP,MHS,BBL}), the
second one relates to $q_1+q_2=0$ and $q_3=0$ \cite{BBL}.

\subsection{Non-minimal Dirac monopoles}\noindent

Here we assume, that electric charge is absent, $Q_{(e)}=0$. One
can check directly, that the equations (\ref{F}) and (\ref{HikR})
are satisfied identically, when the potential of the
electromagnetic field $A_i$ and the field strength tensor $F_{ik}$
outside a point-like magnetic charge $Q_{(m)}$ have the form
\begin{equation}\label{1}
A_k= \frac{Q_{(m)}}{\sqrt{4\pi}}(1-\cos\theta)\delta_k^\varphi\,,
\end{equation}
\begin{equation}\label{2}
F_{ik}= \frac{Q_{(m)}}{\sqrt{4\pi}} \sin\theta
\left(\delta_i^{\theta} \delta_k^{\varphi} - \delta_k^{\theta}
\delta_i^{\varphi} \right) \,.
\end{equation}
Surprisingly, these quantities depend neither on the radial
variable $r$, nor on the coupling parameters $q_1$, $q_2$, $q_3$.
Thus, the well-known solution with a monopole-type magnetic field
satisfies the non-minimally extended Maxwell equations. As a next
step, we solve the Einstein equations, which can be reduced for
the given ansatz to the following pair of key equations
\begin{equation}\label{EinR1}
\frac{\sigma^{\prime}}{\sigma} \left(1- \frac{\kappa q_1}{r^4}
\right) = \frac{\kappa}{r^5} (10q_1+ 4q_2 +q_3) \,,
\end{equation}
\begin{equation}\label{EinR2}
rN^{\prime} \left(1- \frac{\kappa q_1}{r^4} \right) + N \left[ 1 +
\frac{\kappa}{r^4}(13q_1+ 4q_2 +q_3)\right] = 1 -
\frac{\kappa}{2r^2} + \frac{\kappa}{r^4} (q_1+ q_2 +q_3) \,.
\end{equation}
When $q_1\neq0$, these key equations give the following
three-parameter family of solutions
\begin{equation}\label{si2}
\sigma=\left(1-\frac{\kappa q_1}{r^4}\right)^{\beta}\,, \quad
\beta \equiv \frac{10q_1+4q_2+q_3}{4q_1} \,,
\end{equation}
\begin{equation}
N=1-\frac{2M}{r} \left(1{-}\frac{\kappa
q_1}{r^4}\right)^{-(\beta+1)} + \frac{\kappa}{2r}
\int\limits_r^{\infty}\frac{dx}{x^2}\left[1+\frac{6}{x^2}(4q_1+q_2)\right]\left(1-\frac{\kappa
q_1}{x^4}\right)^{\beta} \left(1-\frac{\kappa
q_1}{r^4}\right)^{-(\beta+1)} \,. \label{N2}
\end{equation}
In the special case, when $q_1=0$, the two-parameter family of
solutions takes the form
\begin{equation}\label{si1}
\sigma = \exp\left[-\frac{\kappa(4q_2+q_3)}{4r^4} \right] \,,
\end{equation}
\begin{equation}
N=1-\frac{2M}{r}
\exp\left[{\frac{\kappa(4q_2+q_3)}{4r^4}}\right]
+\frac{\kappa}{2r}
\int\limits_r^{\infty}\frac{dx}{x^2}\left(1+\frac{6q_2}{x^2}\right)
\exp\left[{\frac{\kappa(4q_2+q_3)}{4}\left(\frac{1}{r^4}-\frac{1}{x^4}
\right) }\right]\,.\label{N1}
\end{equation}
Here $\kappa$ is a convenient positive constant with the
dimensionality of area, $\kappa= 2 Q^2_{(m)}$, and $M$ is a
constant of integration describing the asymptotic mass of the
monopole. These solutions are direct $U(1)$-analogs of the
non-minimal Wu-Yang monopole solutions obtained in \cite{BaZaPLB},
and they may be indicated as the non-minimal Dirac monopoles.
Clearly, when $q_1=q_2=q_3=0$, the obtained solutions reduce to
the minimal one (\ref{RN}) with $Q_{(e)}=0$.

\section{Conditions for the absence of naked singularity }\noindent\label{numbersection}

In the papers \cite{BaZaPLB,GraCos08} we attracted a special
attention to the solution (\ref{si2})-(\ref{N1}) with regular
metric. In particular, it was shown that, when $q_1=-q$, $q_2=4q$,
$q_3=-6q$ and $q$ is positive, there are no horizons if the mass
of the monopole is less than some critical mass $M_{(crit)}$. Now
we focus on the analysis of the metrics, which have at least one
horizon for {\it arbitrary} mass and magnetic charge $Q_{(m)}$,
and we search for the relevant relationships between the coupling
constants $q_1$, $q_2$, $q_3$. It is convenient to divide our
analysis into three parts for the cases $q_1 < 0$, $q_1=0$ and
$q_1>0$, respectively.

\subsection{First case: $q_1 < 0$}\noindent

The main problem we are going to solve is the following:  for what
values of $q_1$, $q_2$, $q_3$ the equation
\begin{equation}\label{Nh}
N(r)=0
\end{equation}
has at least one positive solution, when the parameters $M\geq 0$
and $\kappa>0$ are arbitrary. For the derivation of basic
inequalities we use the following method. First, taking into
account (\ref{N2}) we rewrite the equation (\ref{Nh}) in the form
\begin{equation}
2M = r\left(1+\frac{\kappa
    |q_1|}{r^4}\right)^{\beta+1}+\frac{\kappa}{2}\int\limits_r^\infty
    \frac{dx}{x^2}\left[1+\frac{6}{x^2}(4q_1+q_2)\right]\left(1+\frac{\kappa|q_1|}{x^4}\right)^{\beta}.
\end{equation}
Second, we make the replacement $z=r(\kappa|q_1|)^{-1/4}$ in this
equation, thus introducing a new dimensionless variable $z$.
Third, we rewrite the obtained equation as follows
\begin{equation}\label{Sh1}
\frac12\sqrt{\frac{\kappa}{|q_1|}}=S(z) \,,
\end{equation}
\begin{equation}\label{Sh212}
S(z)\equiv \frac{1}{\int_z^\infty
    d\tau\,\tau^{{-}2}\left(1{+}\tau^{{-}4}\right)^\beta}\left[\left(12{-}\frac{3q_2}{|q_1|}\right) \int_z^\infty
    d\tau\,\tau^{{-}4}\left(1{+}\tau^{{-}4}\right)^\beta
+ \frac{2M}{(\kappa|q_1|)^{1/4}}
    - z\left(1{+}z^{{-}4}\right)^{\beta{+}1}
    \right]\,.
\end{equation}
Since for negative $q_1$ the  expression $(1{+}\tau^{{-}4})$,
obtained by replacement, does not take on a zero value, then the
function $S(z)$ is continuous in the interval $z\in(0;+\infty)$.
At the limiting case $z\to +\infty$ this function takes on the
negative infinite value, $\lim\limits_{z\to +\infty}S(z)=-\infty$.
We assume that the equality (\ref{Sh1}) should be fulfilled for
arbitrary magnetic charge, i.e., for arbitrary non-negative value
of the parameter $\sqrt{\frac{\kappa}{|q_1|}}$. Thus, the function
$S(z)$ should reach infinite value at least in one of the points
of the interval $z \in(0,+\infty)$. Being continuous at $z>0$, the
function $S(z)$ can reach infinity only at $z=0$. Consequently,
one should estimate the behaviour of $S(z)$ in the vicinity of
this point. The simple analysis shows that in this limit $S(z)$
tends to infinity, when $\beta \geq -3/4$. In addition, the
infinite value is positive, i.e.,  $S(0)= + \infty $, when
$12-\frac{3q_2}{|q_1|}>4\beta+3$ only. After the substitution of
the expression for $\beta$ from (\ref{si2}) we obtain the basic
inequalities
\begin{equation}\label{maincond}
    13q_1+4q_2+q_3\leq 0\,, \qquad q_1+q_2+q_3>0\,.
\end{equation}

\subsection{Second case: $q_1=0$}

When $q_1$ vanishes we take the equation (\ref{N1}) instead of
(\ref{N2}), and exponential function $\exp\{ \kappa
(4q_2+q_3)/4r^4\}$ instead of $(1+ \kappa|q_1|/r^4)^{\beta}$. The
procedure for obtaining the basic inequalities is  similar to the
one used in the previous case, and it yields the same inequalities
(\ref{maincond}).

\subsection{Third case: $q_1>0$}\noindent

When $q_1$ is positive, the situation differs essentially from
that of two previous cases. First of all,  the metric
(\ref{metrica}), (\ref{si2}), (\ref{N2}) is ill-defined for a
fractional $\beta$, when $r<\sqrt[4]{\kappa q_1}$. If $\beta$ is
an integer, the metric has a singularity at $r=\sqrt[4]{\kappa
q_1}$. Therefore, we have to restrict our consideration by the
interval $r>\sqrt[4]{\kappa q_1}$ only. Let us show now that no
horizon for arbitrary mass and magnetic charge exists for this
interval. The procedure of finding of basic inequalities is
similar to that of the first case, but now we obtain a modified
auxiliary function $\tilde{S}(z)$ instead of $S(z)$ (see
(\ref{Sh212}))
\begin{equation}\label{Sh2122}
\tilde{S}(z)\equiv - \frac{1}{\int_z^\infty
    d\tau\,\tau^{{-}2}\left(1-\tau^{{-}4}\right)^\beta}
    \left[\left(12+\frac{3q_2}{q_1}\right) \int_z^\infty
    d\tau\,\tau^{{-}4}\left(1-\tau^{{-}4}\right)^\beta
- \frac{2M}{(\kappa q_1)^{1/4}}
    + z\left(1-z^{{-}4}\right)^{\beta+1}
    \right]\,.
\end{equation}
The function $\tilde{S}(z)$ is continuous in the interval $z
\in(1;+\infty)$ and $\lim\limits_{z\to
+\infty}\tilde{S}(z)=-\infty$. In order to resolve the equation
\begin{equation}\label{Sh111}
\frac12\sqrt{\frac{\kappa}{q_1}}=\tilde{S}(z) \,,
\end{equation}
for arbitrary magnetic charge, we should require, that
$\tilde{S}(z)$ tends to positive infinity at $z \to 1$, i.e.,
$\lim\limits_{z\to 1}\tilde{S}(z)=+\infty$. However,
$\tilde{S}(1)$ is finite, thus, it is impossible.

\subsection{Basic inequalities}

Summing up the results of three previous subsections, we can
resume, that in the non-minimal model under consideration the
metric (\ref{metrica}), (\ref{si2})-(\ref{N1}) has at least one
event horizon for arbitrary values of the mass $M\geq0$ and magnetic
charge $Q_{(m)}$, when three following inequalities are valid
\begin{equation}\label{mainineq}
     q_1\leq0\,, \qquad 13q_1+4q_2+q_3\leq0\,, \qquad q_1+q_2+q_3>0\,.
\end{equation}
Since the first and second inequalities are unstrict, there are
three interesting particular cases.

\subsubsection{$13q_1+4q_2+q_3 \neq 0$}

If the second inequality is strict, i.e., $\beta\neq-3/4$, the
value of the function $N(r)$ at the center is finite and negative
\begin{equation}\label{0ineq}
N(0)= \frac{(q_1+q_2+q_3)}{(13q_1+4q_2+q_3)} <0 \,.
\end{equation}
Since $N(\infty)=1 >0$, and $N(r)$ is continuous function there is
at least one point at $r>0$, say $r^{*}$, in which $N(r^{*})=0$.
This fact demonstrates explicitly, that the singular point of
origin $r=0$ is hidden inside of an event horizon.

\subsubsection{$13q_1+4q_2+q_3 =0$ and $q_1 \neq 0$ }

When $\beta=-3/4$ and $q_1\neq 0$, the function $N(r)$ behaves in
the vicinity of $r=0$ as
\begin{equation}
N(r)\sim A\ln r\,, \quad  A=\frac{3(4q_1+q_2)}{q_1}>0 \,.
\end{equation}
Thus, at the point of origin $N(0)= -\infty$, and one has at least
one solution of the equation $N(r)=0$, as in the previous case.

\subsubsection{$13q_1+4q_2+q_3 =0$ and $q_1 = 0$}

When $\beta=-3/4$ and $q_1 = 0$, one obtains, that $q_2$ is
negative and at $r \to 0$ the function $N(r)$ behaves as
\begin{equation}
N(r)\sim - \frac{\kappa |q_2|}{r^4} \,.
\end{equation}
Thus, the values $N(0)$ are now infinite, but also negative,
confirming our conclusion, that there exists at least one point
with $N(r^{*})=0$.

The inequalities (\ref{mainineq}) can be rewritten in the simple
form using the following re-parametrization
\begin{equation}\label{reparamet}
     q_1 = - Q_1\,, \qquad q_2 = 4 Q_1 -Q_2 -Q_3\,, \qquad q_3 = -3 Q_1 +Q_2 +4Q_3\,.
\end{equation}
In these new terms the basic inequalities read
\begin{equation}\label{1reparamet}
 Q_1 \geq 0\,, \qquad Q_2 \geq 0\,, \qquad Q_3>0\,,
\end{equation}
separating the first octant with two boundary planes in the
auxiliary three-dimensional space of parameters $Q_1,Q_2,Q_3$. A
true number of horizons for each set of $q_1$, $q_2$, $q_3$,
satisfying (\ref{mainineq}), depends on relations between the
mass, charge, and coupling constants. Below we consider a number
of exact solutions illustrating our conclusions.

\section{Explicit examples of exact solutions with non-minimal horizons}

\subsection{Non-minimal solution of the Reissner-Nordstr\"om
type with $q_1 = 0$, $4q_2+q_3=0$}\noindent\label{RNsection}

The given set of parameters relates to the third (special) case,
considered in the previous subsection.  When $q_1$ vanishes and
$q_3=-4q_2$, the formulas (\ref{si1}) and (\ref{N1}) yield
\begin{equation}
\sigma(r)=1\,,\qquad N(r)=1-\frac{2M}{r} + \frac{\kappa}{2r^2} + \frac{\kappa q_2}{r^4}
\,.\label{N11}
\end{equation}
We deal with the one-parameter non-minimal generalization of the
Reissner-Nordstr\"om solution. This exact solution is
characterized by the infinite central value $N(0)$, this value
being negative if $q_2<0$. Thus, starting from $N(\infty)=1>0$ the
continuous function $N(r)$ tends to $N(0) =- \infty$ and crosses
the line $N=0$ at least once {\it for arbitrary mass and charge}.
In other words, the equation $N(r)=0$ leads to the quartic
equation
\begin{equation}
r^4 - 2M r^3 + \frac{\kappa}{2}r^2 + \kappa q_2=0 \,, \label{N111}
\end{equation}
which has at least one positive real root, and, thus, guarantees
that the space-time possesses at least one horizon for arbitrary
mass and charge. For this case the inequalities (\ref{mainineq})
yields  that $-3q_2>0$, in agreement with our conclusion.

Generic requirements for $M$, $\kappa$ and $q_2$, which classify
the number of non-minimal horizons, can be found using the
well-known Ferrari method (see, e.g., \cite{Ferrari}),
nevertheless, we restrict ourselves by two explicit examples only,
demonstrating the cases with one and three horizons.

\subsubsection{$M=0$: One horizon}

In the minimal model the condition $M=0$ leads to the
Reissner-Nordstr\"om solution with naked singularity. In the
non-minimal model the quartic equation (\ref{N111}) reduces to the
biquadratic one, and, clearly, the only positive real root is
\begin{equation}
r= r_{({\rm H})}=\frac{1}{2} \sqrt{\kappa}\sqrt{\sqrt{1+\frac{16
|q_2|}{\kappa}}-1} \,. \label{N112}
\end{equation}
In the minimal limit  $q_2 \to 0$ the radius of the horizon
$r_{({\rm H})}$ tends to zero. When $|q_2| \ll \kappa$, $r_{({\rm
H})} \to \sqrt{2|q_2|}$; when $|q_2| \gg \kappa$, $r_{({\rm H})}
\to (\kappa|q_2|)^{\frac{1}{4}}$.

\subsubsection{$\kappa = 2M^2$: Three horizons}

In the minimal model the condition $\kappa = 2M^2$ (or
equivalently, $M^2=Q^2_{(m)}$) introduces the so-called extreme
Reissner-Nordstr\"om black hole, for which two horizons coincide.
For the non-minimal model the equation (\ref{N111}) can be
presented as a product of two quadratic equations. Clearly, for
arbitrary mass there exists the positive real root
\begin{equation}
r_{({\rm H}1)}= \frac{M}{2}\left(1+\sqrt{1+\frac{4
\sqrt{2|q_2|}}{M}} \right) \,. \label{N113}
\end{equation}
In addition, when $M>4\sqrt{2|q_2|}$, there are two roots else
\begin{equation}
r_{({\rm H}2,3)}= \frac{M}{2}\left(1 \pm \sqrt{1-\frac{4
\sqrt{2|q_2|}}{M}} \right)\,. \label{N114}
\end{equation}
When $q_2 \to 0$, one obtains from (\ref{N113}) and (\ref{N114})
\begin{equation}
r_{({\rm H}1)} \simeq M + \sqrt{2|q_2|} \,, \quad r_{({\rm H}2)}
\simeq M - \sqrt{2|q_2|} \,, \quad r_{({\rm H}3)} \simeq
\sqrt{2|q_2|} \,. \label{N01}
\end{equation}
This means that non-minimal coupling removes the degeneration,
which appears if the mass coincides with the charge, and splits
the double horizon of the extreme Reissner-Nordstr\"om magnetic
black hole into two space-apart horizons with the radii $r_{({\rm
H}1)}$ and $r_{({\rm H}2)}$, respectively. The radius of the third
non-minimal horizon $r_{({\rm H}3)}$ tends to zero at vanishing
coupling parameter $q_2$.

\subsection{Non-minimal model of the Drummond-Hathrell type}\noindent\label{DHsection}

The one-parameter Drummond-Hathrell model  arises from the
calculation of the one-loop QED-corrections to the
Einstein-Maxwell Lagrangian in curved space-time \cite{Drum}. For
this model $q_1=-5q$, $q_2=13q$, $q_3=-2q$, where
$q=\frac{\alpha\lambda^2}{180\pi}$ ($\alpha\approx 1/137$ is the
fine structure constant, $\lambda\approx 4\cdot 10^{-13}\
\mathrm{m}$ is the Compton wavelength of the electron). Clearly,
\begin{equation}\label{DH1}
q_1\leq0\,, \quad 13q_1+4q_2+q_3 = - 15 q < 0\,, \quad q_1+q_2+q_3
= 6q > 0\,,
\end{equation}
i.e.,  this set of the coupling constants satisfies basic
inequalities (\ref{mainineq}).

\subsubsection{Number of horizons}\noindent

In the Drummond-Hathrell model $\beta=0$, and the metric functions
$\sigma(r)$ and $N(r)$ take the following explicit form
\cite{Zayatsthesis}
\begin{equation}\label{sNDH}
    \sigma(r)=1\,,\quad N(r)=\frac{r^4-2Mr^3+\kappa r^2/2-2\kappa q}{r^4+5\kappa q}\,.
\end{equation}
At the point of origin $N(0) = -2/5 < 0$ in agreement with
(\ref{0ineq}), as well as $N(\infty)=1$, thus, at least one
horizon exists for arbitrary mass and charge. Let us mention that
$N(0) \neq 1$, consequently, the metric (\ref{sNDH}) possesses the
so-called ``mild'' or ``conic'' singularity. This means that the
metric functions themselves, $\sigma(r)$ and $N(r)$, are finite at
$r=0$, whereas the Ricci scalar is infinite because of the term
$[1-N(r)]/r^2$. The same situation is described in \cite{BBL} for
the Fibonacci model.

In order to find the number of horizons for the metric
(\ref{sNDH}), let us consider in more detail the roots of the
numerator of $N(r)$, i.e., analyse the quartic equation
\begin{equation}\label{f}
r^4-2Mr^3+ \frac{\kappa r^2}{2}-2\kappa q=0\,.
\end{equation}
We divide the analysis into two cases: $\kappa>96q$ and
$\kappa\leq 96q$, respectively. When $\kappa>96q$, it is
convenient to introduce the following auxiliary quantities
\begin{equation}
M_{1,2}=\frac{2r_{1,2}}{3}+\frac{\kappa}{6 r_{1,2}}\,, \quad
r_{1,2}=\frac{\sqrt\kappa}{2}\cdot\left(1\pm\sqrt{1-\frac{96q}{\kappa}}\right)^{1/2}\,.
\end{equation}
There are three different possibilities:

\noindent {\it (i)} $M_1 <M < M_2$ : Equation (\ref{f}) has three
real positive solutions;

\noindent {\it (ii)} $M= M_1$ or $M = M_2$ : There are two
different solutions, since a couple of solutions coincide;

\noindent {\it (iii)} $M < M_1$ or $M > M_2$: Equation (\ref{f})
has only one real positive solution.

\noindent When $\kappa\leq 96q$, the equation (\ref{f}) has only
one positive real root for arbitrary mass $M$. In other words, for
arbitrary magnetic charge (i.e., for any $\kappa$) one can find at
least one horizon attributed to any mass $M$, the naked
singularity does not exist in the non-minimal Drummond-Hathrell
model.

As a simple explicit illustration let us assume that the monopole
mass $M$ is vanishing. Then the single positive solution to
(\ref{f}) can be written in the explicit form
\begin{equation}\label{rh0}
 r_{h0}=\frac{\sqrt\kappa}{2}\cdot\left(\sqrt{1+\frac{32q}{\kappa}}-1\right)^{1/2}\,.
\end{equation}
If $q=0$, this horizon turns into the point of origin. When
$q\ll\kappa$, $r_{h0}$ tends to $2\sqrt{q}$, when $q\gg\kappa$,
$r_{h0}\approx\sqrt[4]{2\kappa q}$. Thus, this horizon is
essentially non-minimal.

\subsubsection{Numerical estimation of the radius of the non-minimal horizon}

The non-minimal Drummond-Hathrell model is especially attractive,
since all the parameters of the model can be directly estimated.
Indeed, the value of $q$ can be readily estimated as
$q=\frac{\alpha\lambda^2}{180\pi}\approx 2\cdot 10^{-30}\
\mathrm{m}^2$.  The quantity $\sqrt{\kappa}$ is proportional to
the magnetic charge $Q_{(m)}$, and for a magnetic monopole with
unit charge it can be estimated as $\sqrt{\kappa} \simeq 10^{-34}\
\mathrm{m}$ \cite{BaSuZa}. Thus, we deal with the case
$q\gg\kappa$, the inequality $\kappa \leq 96q$ is valid, and there
is only one horizon according to our previous analysis. The radius
of non-minimal horizon can be found now from the formula
\begin{equation}\label{rH}
   r_h\approx(2\kappa q)^{1/4}\sim 10^{-25} \ \mathrm{m}\,.
\end{equation}
The choice of this formula can be motivated as follows. The mass
of monopole is unknown, but we assume, that it is less than the
Planck mass, which guarantees that $M\ll\sqrt\kappa$. Then, using
the formula (\ref{rh0}) for vanishing mass, and taking into
account that $q\gg\kappa$, we obtain (\ref{rH}). Thus, our
conclusion is that the non-minimal horizon in the
Drummond-Hathrell model has the radius of the order $10^{-25} \
\mathrm{m}$. This value is much greater than the Planck length,
$L_{\rm pl}\sim 10^{-35}\ {\rm m}$, but is much smaller than the
Compton wavelength of the electron $\lambda \approx 4 \cdot
10^{-13}\ \mathrm{m}$.

\section{Discussion}\noindent

The logic of the development of the non-minimal Einstein-Maxwell
theory prompts, that the phenomenologically introduced coupling
constants $q_1$, $q_2$ and $q_3$, which have the dimensionality of
area, either have to be associated with some known constants of
Nature, or some new non-minimal radii should be introduced and
properly motivated. One attempt to realize this idea was made in
\cite{BL05}, where the approach based on the symmetry of the
susceptibility tensor ${\cal R}^{ikmn}$ (\ref{sus}) is proposed.
In \cite{BBL,BaZaPLB,GraCos08,BZ05} special sets of coupling
parameters were found, for which the metric functions of
non-minimally coupled systems happened to be regular, and the
absence of singularity became one of the arguments for the
non-minimal extension of the Einstein-Maxwell theory.

Here we analysed a new possibility to fix the coupling constants,
which is related to the censorship conjecture. We discussed the
three-parameter family of exact solutions of the non-minimal
Einstein-Maxwell model, which can be associated with magnetic
monopoles of the Dirac type. We have shown explicitly, that the
singular point $r=0$ appears to be hidden by some non-minimal
horizon independently on the mass and magnetic charge, when the
basic inequalities (\ref{mainineq}) are satisfied. In terms of new
appropriate parameters $Q_1$, $Q_2$ and $Q_3$ (see
(\ref{reparamet})) such kind of non-minimal clothing is possible,
when these new parameters belong to the first octant of the
auxiliary three-dimensional ${\bf Q}$-space (including two of
three separating planes). As it was shown by the example of the
non-minimal Drummond-Hathrell model (see
Subsection~\ref{DHsection}), the radius of the non-minimal event
horizon can be estimated  as $r_h\sim 10^{-25}\ \mathrm{m}$, i.e.,
it can be greater by ten orders than the Planck length $L_{\rm
pl}$. In forthcoming papers we intend to analyse  non-minimal
models with electric charge and the dyonic model in order to find
analogous necessary conditions prescribed by the censorship
conjecture. We believe that the combination of requirements
obtained for non-minimal magnetic monopoles, electrically charged
objects and dyons could fix the choice of coupling constants and
define unambiguously the radius of the event horizon, $r_{(P)}$,
associated with the censorship conjecture, proposed by Penrose.

\appendix

\section*{Acknowledgments}\noindent

This work was supported by the Deutsche Forschungsgemeinschaft
through the project No. 436RUS113/487/0-5 and partially by the
Russian Foundation for Basic Research through the grant No.
08-02-00325-a.

\end{document}